\documentstyle[aps,prl,preprint,floats,epsfig]{revtex}  

\begin{document}

\tighten 
\draft 
\preprint{\tighten\vbox{\hbox{CLNS 97-1522 \hfill}
                        \hbox{CLEO 97-27   \hfill}}}

\title{Observation of Exclusive Two-body B Decays to Kaons and Pions}

\date{\today}

\maketitle

\begin{abstract}
\tighten
We have studied two-body charmless hadronic
decays of $B$ mesons into the final states
$\pi\pi$, $K \pi$, and $KK$.
Using 3.3~million
$B\bar{B}$~pairs 
collected with the CLEO-II detector,
we have made the first observation of the decays $B^0\to K^+\pi^-$, 
$B^+\to K^0\pi^+$, and 
the sum of $B^+ \rightarrow \pi^+\pi^0$ and $B^+ \rightarrow K^+\pi^0$
decays (an average over charge-conjugate states is always implied).
We place upper limits on 
branching fractions for the remaining decay modes.
\end{abstract}

\pacs{PACS numbers:13.25.Hw,14.40.Nd}

\begin{center}
R.~Godang,$^{1}$ K.~Kinoshita,$^{1}$ I.~C.~Lai,$^{1}$
P.~Pomianowski,$^{1}$ S.~Schrenk,$^{1}$
G.~Bonvicini,$^{2}$ D.~Cinabro,$^{2}$ R.~Greene,$^{2}$
L.~P.~Perera,$^{2}$ G.~J.~Zhou,$^{2}$
% ...
\end{center}
 
\small
\begin{center}
$^{1}${Virginia Polytechnic Institute and State University,
Blacksburg, Virginia 24061}\\
$^{2}${Wayne State University, Detroit, Michigan 48202}\\
% ...
$^{13}${McGill University, Montr\'eal, Qu\'ebec, Canada H3A 2T8 \\
and the Institute of Particle Physics, Canada}\\
$^{14}${Ithaca College, Ithaca, New York 14850}\\
% ...
$^{25}${Vanderbilt University, Nashville, Tennessee 37235}
\end{center}

  The phenomenon of $CP$ violation, so far observed only in 
the neutral kaon system, can be accommodated  by a complex phase in the   
Cabibbo-Kobayashi-Maskawa (CKM) quark-mixing matrix~\cite{CKM}.
Whether this phase is the correct, or only, source of $CP$ violation
awaits experimental confirmation.
$B$ meson decays, in particular charmless $B$ meson decays,
will play an important role in verifying this picture.

\begin{figure}
\begin{center}
\epsfig{file=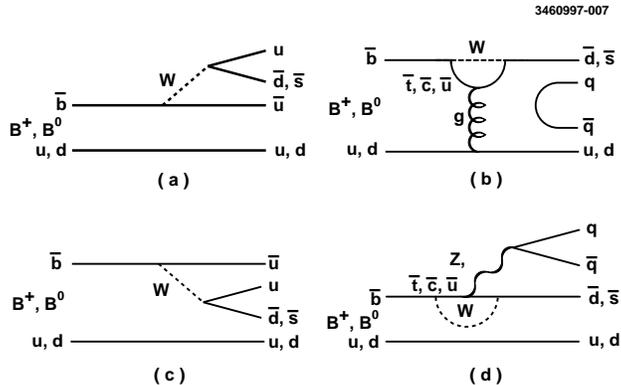, width=3.25in, clip=}
\end{center}
\caption{The dominant decay processes are expected to be 
(a) external W-emission, (b) gluonic penguin, (c) internal W-emission,
(d) external electroweak penguin.}
\label{fig:feynman}
\end{figure}

The decay $B^0 \rightarrow \pi^+\pi^-$, dominated by the $b \rightarrow
u$ tree diagram (Fig.~\ref{fig:feynman}(a)), can be used to measure $CP$ 
violation due to $B^0-\bar B^0$\ mixing at both asymmetric
$B$~factories and hadron colliders. 
However, theoretical uncertainties due to the presence of 
the $b\to dg$ penguin diagram (Fig.~\ref{fig:feynman}(b)) make it
difficult to extract the angle $\alpha$ of the unitarity triangle from 
$B^0 \rightarrow \pi^+\pi^-$ alone. 
Additional measurements of 
$B^+ \rightarrow \pi^+\pi^0$, $B^0 \rightarrow \pi^0\pi^0$, and the use
of isospin 
symmetry may resolve these uncertainties~\cite{isospin}.

 $B\to K\pi $\ decays are dominated by the
$b \rightarrow sg$ gluonic penguin diagram,
with additional contributions from 
$b \rightarrow u$ tree and color-allowed 
electroweak penguin (Fig.~\ref{fig:feynman}(d)) processes.
Interference between the penguin and spectator amplitudes 
can lead to direct $CP$ violation, which
 would manifest itself as a rate asymmetry for
decays of $B$ and $\bar{B}$~mesons.
Recently, the ratio
 $R={\cal B}(B\to K^{\pm}\pi^{\mp})/{\cal B}(B^{\pm}\to K^0\pi^{\pm}) $, 
was shown~\cite{fleischermannel} to constrain $\gamma$, the phase of $V_{ub}$. 
Several methods of measuring $\gamma$ 
using only decay rates of $B\to K\pi,~\pi\pi$ processes were also 
proposed~\cite{triangles}. 
This is particularly important, as $\gamma $\ is the
least known parameter of the 
unitarity triangle and is likely to remain
the most difficult to determine experimentally. This Letter describes
the first measurement of exclusive charmless hadronic $B$ decays. 
Previous measurements existed only 
for the sum of several two-body final states~\cite{bigrare,lepresults}.

The data set used in this analysis was collected with the CLEO-II
detector~\cite{detector} at the Cornell Electron Storage Ring (CESR).
It consists of $3.14~{\rm fb}^{-1}$ taken at the $\Upsilon$(4S)
(on-resonance) and $1.62~{\rm fb}^{-1}$ taken 
below $B\bar{B}$ threshold.  The
on-resonance sample contains 3.3~million $B\bar{B}$ pairs.
The below-threshold sample
is used for continuum background studies.

Charged tracks are
required to pass track quality cuts based on the average hit residual
and the impact parameters in both the $r-\phi$ and $r-z$ planes.
Pairs of tracks with vertices displaced by at least $3$~mm from the primary
interaction point are taken as $K_S^0$ candidates. 
We require
the $\pi^+\pi^-$ invariant mass to be within $10$~MeV, two standard
deviations ($\sigma$), 
of the $K^0_S$\ mass.
Isolated showers with energies greater than
$30$~MeV in the central region of the CsI calorimeter
and greater than $50$~MeV elsewhere, are defined to be photons.
Pairs of photons with an invariant mass within $20$~MeV ($\sim 2\sigma$)
of the nominal $\pi^0$ mass
are kinematically fitted with the mass constrained to the
$\pi^0$ mass.  To reduce combinatoric backgrounds we
require the
lateral shapes of the showers to be consistent with those from photons, 
and that $|\cos\theta^*|<0.97$, where $\theta^*$ is the angle between the
direction of flight of the $\pi^0$ and the photons in the $\pi^0$ rest
frame.

Charged particles are identified as kaons or pions using $dE/dx$.
Electrons are rejected
based on $dE/dx$ and the ratio of the track momentum to the associated shower
energy in the CsI calorimeter.
We reject muons
by requiring that the tracks do not penetrate the steel absorber to a
depth greater than five nuclear interaction lengths.
We have studied the $dE/dx$\ separation between 
kaons and pions for momenta $p \sim 2.6$~GeV$/c$\ in data using
$D^{*+}$-tagged 
$D^0\rightarrow K^-\pi^+$ decays; we find a separation of $(1.7\pm0.1)\sigma$.

We calculate a beam-constrained $B$ mass 
$M = \sqrt{E_{\rm b}^2 - p_B^2}$, where $p_B$ is the $B$
candidate momentum and $E_{\rm b}$ is the beam energy.
The resolution in $M$\ ranges from 2.5 to 3.0~${\rm MeV}/{\it c}^2$, 
where the 
larger resolution corresponds to decay modes with $\pi^0$'s.
We define $\Delta E = E_1 + E_2 - E_{\rm b}$, where $E_1$ and $E_2$
are the energies of the daughters of the $B$ meson candidate.
The resolution on $\Delta E$ is mode-dependent
and ranges from $\pm 26$~MeV for $K^0_S\pi^+$\ 
to $+82/-162$~MeV for $\pi^0\pi^0$.
The latter resolution is asymmetric because of energy loss out of the
back of the CsI crystals.
The energy constraint also helps to distinguish between modes of
the same topology.  
For example, $\Delta E$
for  $B^0\rightarrow K^+\pi^-$,  calculated assuming 
$B^0\rightarrow\pi^+\pi^-$,
has a distribution that is centered at $-42$~MeV, giving a separation
of $1.6\sigma$ between $B^0 \rightarrow K^+\pi^-$ and
$B^0 \rightarrow \pi^+\pi^-$.
We accept events with $M$\ within $5.2-5.3$~$\rm {GeV/c^2}$\ and 
$|\Delta E|<200(300)$~MeV
for decay modes without (with) a $\pi^0$\ in the final state. This fiducial
region 
includes the signal region, and a sideband for background determination.

We have studied backgrounds from $b\to c$\ decays and other $b\to u$\ and
$b\to s$\ decays and find that all are negligible for the analyses presented
here. The main background arises from $e^+e^-\to q\bar q$\ (where $q=u,d,s,c$).
Such events typically exhibit a two-jet structure and can produce high 
momentum back-to-back tracks in the fiducial region.
To reduce contamination from these events, we calculate the angle $\theta_S$
between the sphericity axis of the candidate tracks and showers and the
sphericity axis of the rest of the event. The distribution of $\cos\theta_S$\ 
is strongly peaked at $\pm1$ for $q\bar q$\ events and is nearly flat 
for $B\bar B$\ 
events. We require $|\cos\theta_S|<0.8$\ which eliminates  $83\%$\ 
of the background. 
Using a detailed GEANT-based Monte-Carlo simulation~\cite{geant}
we determine overall detection efficiencies (${\cal E}$) of $5-44\%$,
 as listed in
Table~\ref{tab:results}. Efficiencies contain branching fractions for
$K^0\to K^0_S\to \pi^+\pi^-$\ and $\pi^0\to \gamma\gamma$ where applicable.
We estimate a systematic error on the efficiency using
independent data samples.

Additional discrimination between signal and $q\bar q$\ background is provided
by a Fisher discriminant technique as described in detail in 
Ref.~\cite{bigrare}.
The Fisher discriminant is a linear combination 
${\cal F}\equiv \sum_{i=1}^{N}\alpha_i y_i$\ where the coefficients 
$\alpha_i$ are chosen to maximize the separation between the signal
and background Monte-Carlo samples. 
The 11 inputs, $y_i$, are $|\cos\theta_{cand}|$ (the cosine of the angle 
between the candidate sphericity axis and beam axis), the
ratio of Fox-Wolfram moments $H_2/H_0$~\cite{fox}, and nine variables that 
measure the scalar sum of the momenta of tracks and showers from the rest of 
the event in 
nine angular bins, each of $10^\circ$, centered about the candidate's
sphericity axis.

\begin{table}
\begin{center}
\caption{Experimental results and theoretical predictions [10]. 
Branching fractions
(${\cal B}$) and 90\% C.L. upper limits are given in $10^{-5}$ units.
Quoted significance of the fit results is statistical only.
The errors on ${\cal B}$ are statistical, fit systematics, and efficiency
systematics respectively.}
\begin {tabular}{l c c c c c}
Mode&   $N_S$ & Sig. & ${\cal E}(\%)$ &${\cal B}$
& Theory ${\cal B}$\\
\hline
$\pi^+\pi^-$ &$9.9^{+6.0}_{-5.1}$  & 2.2$\sigma$        &$44\pm3$
& $<~1.5$  & 0.8--2.6      \\
$\pi^+\pi^0$ & $11.3^{+6.3}_{-5.2}$ & 2.8$\sigma$        &$37\pm3$
& $<~2.0$  & 0.4--2.0        \\
$\pi^0\pi^0$ & $2.7^{+2.7}_{-1.7}$ & 2.4$\sigma$        &$29\pm3$
& $<0.93$ & 0.006--0.1       \\
\hline
$K^+\pi^-$   &$21.6^{+6.8}_{-6.0}$ & 5.6$\sigma$         &$44\pm3$
&$1.5^{+0.5}_{-0.4}\pm0.1\pm 0.1$ & 0.7--2.4        \\
$K^+\pi^0$   &$8.7^{+5.3}_{-4.2}$  & 2.7$\sigma$        &$37\pm3$
& $<~1.6$  & 0.3--1.3         \\
$K^0\pi^+$   &$9.2^{+4.3}_{-3.8}$  & 3.2$\sigma$       &$12\pm1$
&$2.3^{+1.1}_{-1.0}\pm0.3\pm 0.2$ &  0.8--1.5               \\
$K^0\pi^0$   & $4.1^{+3.1}_{-2.4}$ & 2.2$\sigma$        &$8\pm1$
& $<~4.1$  & 0.3--0.8     \\
\hline
$K^+K^-$     &$0.0^{+1.3}_{-0.0}$   & 0.0$\sigma$       &$44\pm3$
& $<~0.43$ &       --              \\
$K^+\bar{K}^0$   & $0.6^{+3.8}_{-0.6}$  & 0.2$\sigma$       &$12\pm1$
& $<~2.1$  & 0.07--0.13   \\
$K^0\bar{K}^0$   & 0   & --      &$5\pm 1$
& $<~1.7$  & 0.07--0.12   \\
\hline
$h^+\pi^0$   &$20.0^{+6.8}_{-5.9}$ & 5.5$\sigma$         &$37\pm3$
&$1.6^{+0.6}_{-0.5}\pm0.3\pm 0.2$  & --         \\
\end {tabular}
\label{tab:results}
\end{center}
\end {table}

For all modes except $B^0\to K^0\bar{K}^0$ we perform unbinned
maximum-likelihood (ML) fits using $\Delta E$, $M$, ${\cal F}$,
 $|\cos\theta_B|$ (the angle between the $B$ meson momentum and beam axis), 
and $dE/dx$ (where applicable) as input information for each candidate
event to determine the signal yields. 
Five different fits are performed, one for each topology 
($h^+h^-$, $h^+\pi^0$, $\pi^0\pi^0$, $h^+K^0_S $, and $K^0_S\pi^0$,
$h^\pm$\ referring to a charged kaon or pion).
In each of these fits the likelihood of the  event
is parameterized by the sum of probabilities for
all relevant signal and background hypotheses,
with relative weights determined by maximizing likelihood function ($\cal L$).
The probability of a particular hypothesis is calculated as a product of the
probability density functions (PDF's) for each of the input variables.
The PDF's of the input variables are parameterized by a Gaussian,
a bifurcated Gaussian, or
a sum of two bifurcated Gaussians, except for $|\cos\theta_B|$ 
($1-|\cos\theta_B|^2$ for 
signal, constant for background),
background 
$\Delta E$ (straight line), and background $M$\ 
($f(M)\propto M\sqrt{1-x^2}\exp[-\gamma(1-x^2)]$; $x=M/E_b$)
~\cite{argusbackground}.

The parameters for the PDF's 
are determined from independent data and high-statistics Monte-Carlo
samples. We estimate a systematic error on the fitted yield by varying the
PDF's used in the fit. The error is dominated by the limited statistics in the
independent data samples we used to determine the PDF's.
Further details about the likelihood fit can be found in Ref.~\cite{bigrare}.

\begin{figure}[htbp]
\begin{center}
\epsfig{file=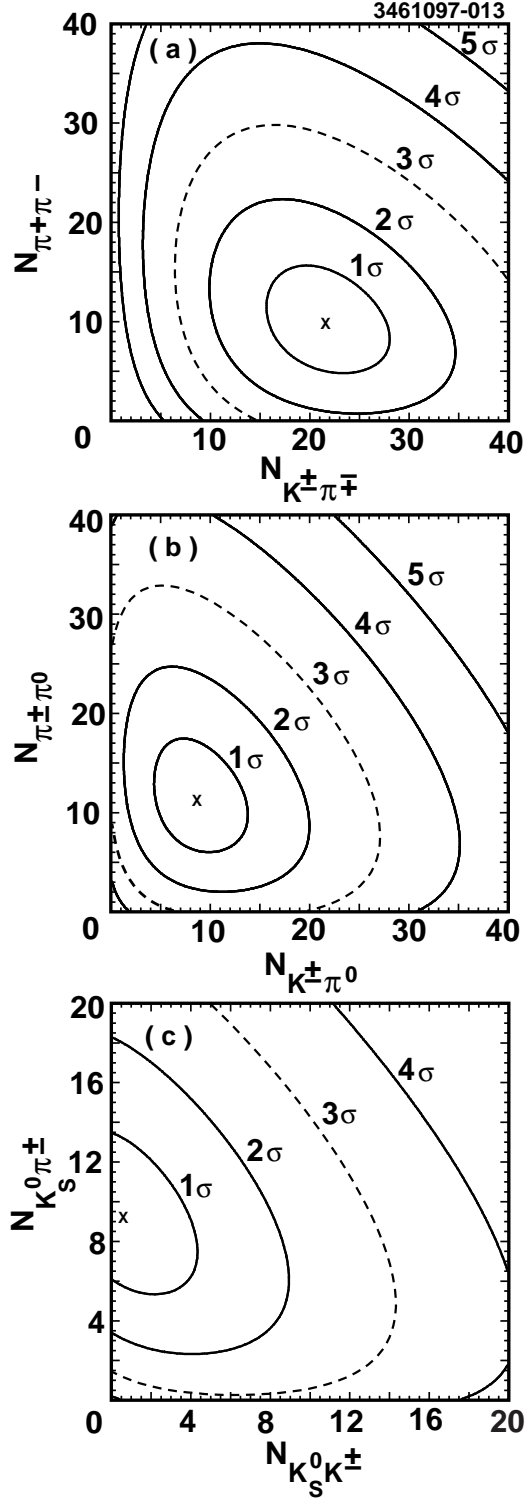,width=2.9in}
\end{center}
\caption{Contours of the $-2\ln{\cal L}$ for the ML fits to 
(a) $N_{K^{\pm}\pi^{\mp}}$ and $N_{\pi^+\pi^-}$ for $B^0\rightarrow K^+\pi^-$
and $B^0\rightarrow \pi^+\pi^-$; 
(b) $N_{K\pi^0}$ and $N_{\pi\pi^0}$ for
$B^+\rightarrow K^+\pi^0$ and $B^+\rightarrow \pi^+\pi^0$;
(c) $N_{K^0_SK}$ and $N_{K^0_S\pi}$ for
$B^+\rightarrow \bar{K}^0K^+$
and $B^+\rightarrow K^0\pi^+$.
}
\label{fig:contourkpi}
\end{figure}

\begin{figure}[htbp]
\begin{center}
\epsfig{file=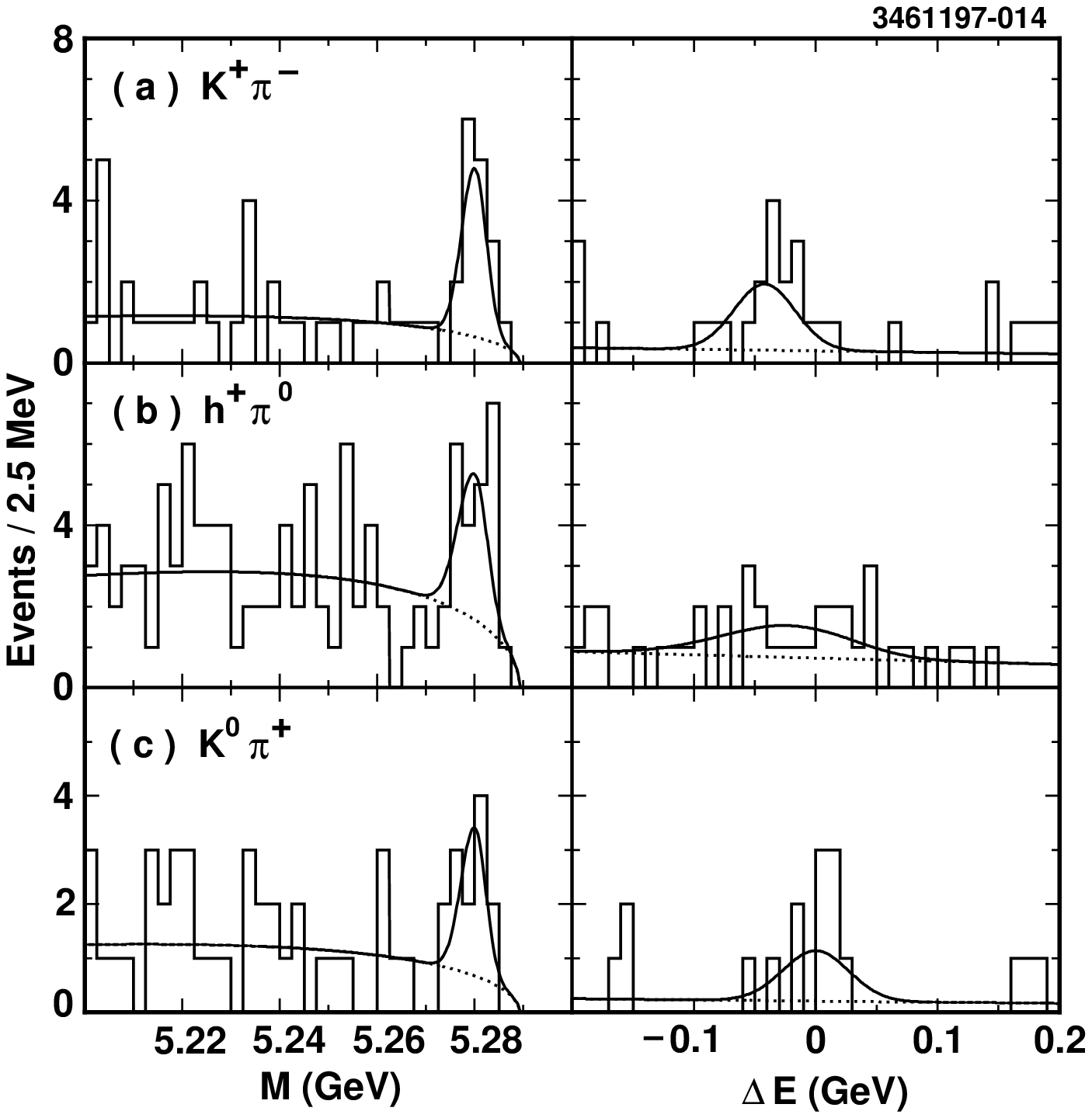,width=3.25in}
\end{center}
\caption{$M$ and $\Delta E$ plots for
(a) $B^0\rightarrow K^+\pi^-$,
(b) $B^+\rightarrow h^+\pi^0$, and (c) $B^+\rightarrow K^0\pi^+$.
The scaled projection of the total likelihood fit (solid curve)
and the continuum background component (dotted curve) are overlaid.}
\label{fig:mass_2body}
\end{figure}

Figure \ref{fig:contourkpi}  shows
contour plots of $-2\ln{\cal L}$ for the ML fits to the signal yields 
($N$). 
The curves represent 
the $n\sigma$ contours ($n=1-5$), which correspond to the increase
 in $-2\ln{\cal L}$ by $n^2$. The dashed curve marks the $3\sigma$ contour.
The statistical significance of a given signal yield is determined by
repeating the fit with the signal yield fixed to be zero and recording 
the change in $-2\ln{\cal L}$.
  To further illustrate the fits,
Fig.~\ref{fig:mass_2body} shows $M$ ($\Delta E$) projections for events
in a signal region defined by
$|\Delta E| < 2\sigma_{\Delta E}$ ( $|M-5.28| < 2\sigma_{M}$).
 We also make a cut on ${\cal F}$ which keeps
$67\%$ of the signal and rejects $80\%$ of the background. 
For Fig.~\ref{fig:mass_2body}(a), events are 
sorted by 
$dE/dx$ according to the most likely hypothesis.
For Fig.~\ref{fig:mass_2body}(c), $3\sigma$ consistency with the
pion hypothesis is required.
Overlaid on these plots are the projections of the
PDF's used in the fit, normalized according to the fit results multiplied by 
the efficiency of the additional cuts ($\sim 60-70\%$ for the signal 
and $\sim 2-10\%$ for the background).
The central values of the signal yields
from the fits ($N_S$) are given in
Table~\ref{tab:results}. We find statistically significant signals 
for the decays $B^0\to K^+\pi^-$ 
and $B^+\to K^0\pi^+$.
The latter mode constitutes the first unambiguous observation of a
gluonic penguin decay. 
The former mode may have a sizeable contribution from the color-allowed
$b \rightarrow u$ tree-level spectator diagram in addition to the dominant
gluonic penguin amplitude. 
We also observe a significant signal in the sum of 
decays  $B^+\to K^+\pi^0$ and $B^+\to \pi^+\pi^0$.

As a cross-check, we perform a counting analysis in the modes
$B^0\rightarrow K^+\pi^-,~B^+\to K^0\pi^+$, and
$B^+\to h^+\pi^0$.
We calculate the probability of the background fluctuation to produce 
the excess of events shown in Fig. \ref{fig:mass_2body}
 to be $2.0\times 10^{-7}$ for the $K^+\pi^-$
mode, $1.6\times 10^{-3}$ for the $h^+\pi^0$ mode, and $2.5\times 10^{-4}$ for the $K^0\pi^+$ mode.

The statistical significance of the fitted yields in the modes 
$\pi^+\pi^-,\ \pi^+\pi^0,\ \pi^0\pi^0,\ K^+\pi^0$,\ and $K^0\pi^0$\ ranges
 from 
$2.2\sigma$ to $2.8\sigma$. We consider these to be not statistically 
significant and
calculate $90\%$\ confidence level (C.L.) upper limit yields by
integrating the likelihood function
\begin{equation}
{\int_0^{N^{UL}} {\cal L}_{\rm max} (N) dN
\over
\int_0^{\infty} {\cal L}_{\rm max} (N) dN}
= 0.90
\nonumber
\end{equation}
where ${\cal L}_{\rm max}(N)$ is the maximum $\cal L$\ at fixed $N$\ to
conservatively account for possible correlations among
the free parameters in the fit. We then increase upper limit yields by
their systematic errors and reduce
detection efficiencies by their systematic errors to calculate 
branching fraction upper limits given in Table I.

We search for the decay $B^0\to K^0\bar{K}^0$\ via 
$K^0,\bar{K}^0\to K^0_S\to \pi^+\pi^-$.
 Since the background 
for this decay is quite low, the complication of a ML fit is not 
necessary and a simple counting analysis is used. 
Event selection is as described above,
 except no Fisher discriminant is used and $|\cos\theta_T|<0.75$ cut 
is applied ($\cos\theta_T$ is defined similar to $\cos\theta_S$, but with 
thrust axis used instead of sphericity).
We define the signal region by requiring $|\Delta E|<65$~MeV 
($2.5\sigma$), and $|M-5.28| <0.005$~GeV$/c^2$\ ($2.4\sigma$).
We observe no events in the signal region and calculate a $90\%$ C.L.
branching fraction upper limit of  
${\cal B}(B^0\to K^0\bar{K^0}) < 1.7\times 10^{-5}$.

As a comparison, we relate 
$B\to\pi l\nu$ and $B\to\pi\pi$ processes within the factorization
hypothesis. Using the ISGW II~\cite{isgwii} form factors,
the QCD factor $a_1 = 1.03\pm 0.07$~\cite{jorge},
and the CLEO measurement
${\cal B}(B^0\to\pi^-l^+\nu)=(1.8\pm 0.4\pm 0.3\pm 0.2)\times
10^{-4}$~\cite{pilnu}, we predict
${\cal B}(B^0\to\pi^+\pi^-)=(1.2\pm 0.4)\times 10^{-5}$\ and
${\cal B}(B^+\to\pi^+\pi^0)=(0.6\pm 0.2)\times 10^{-5}$~\cite{errors}.
These predictions are consistent with our upper limits as well as central 
values from the fit:
${\cal B}(B^0\to\pi^+\pi^-)=(0.7\pm 0.4)\times 10^{-5}$\ and
${\cal B}(B^+\to\pi^+\pi^0)=(0.9^{+0.6}_{-0.5})\times 10^{-5}$.

In summary, we have measured branching fractions for 
two of the four exclusive $B\to K\pi$ decays,
while only upper limits could be established for the processes 
$B\to \pi\pi, KK$. Our results therefore indicate that the 
$b\to sg$ penguin amplitude dominates charmless hadronic $B$ decays.

We gratefully acknowledge the effort of the CESR staff in providing us with
excellent luminosity and running conditions.
This work was supported by 
the National Science Foundation,
the U.S. Department of Energy, 
the Heisenberg Foundation,  
the Alexander von Humboldt Stiftung,
Research Corporation,
the Natural Sciences and Engineering Research Council of Canada,
and the A.P. Sloan Foundation.


\begin{references}
\bibitem{errors}
The errors quoted do not include theoretical uncertainties due to the
factorization hypothesis or form factors.
\end{references}
\end{document}